# scientific reports

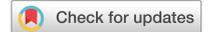

**OPEN**

# Drug repurposing for COVID-19 using graph neural network and harmonizing multiple evidence

Kanglin Hsieh[1], Yinyin Wang[2], Luyao Chen[1], Zhongming Zhao[3], Sean Savitz[4], Xiaoqian Jiang[1], Jing Tang[2] & Yejin Kim[1]✉

Since the 2019 novel coronavirus disease (COVID-19) outbreak in 2019 and the pandemic continues for more than one year, a vast amount of drug research has been conducted and few of them got FDA approval. Our objective is to prioritize repurposable drugs using a pipeline that systematically integrates the interaction between COVID-19 and drugs, deep graph neural networks, and in vitro/ population-based validations. We first collected all available drugs (n = 3635) related to COVID-19 patient treatment through CTDbase. We built a COVID-19 knowledge graph based on the interactions among virus baits, host genes, pathways, drugs, and phenotypes. A deep graph neural network approach was used to derive the candidate drug's representation based on the biological interactions. We prioritized the candidate drugs using clinical trial history, and then validated them with their genetic profiles, in vitro experimental efficacy, and population-based treatment effect. We highlight the top 22 drugs including Azithromycin, Atorvastatin, Aspirin, Acetaminophen, and Albuterol. We further pinpointed drug combinations that may synergistically target COVID-19. In summary, we demonstrated that the integration of extensive interactions, deep neural networks, and multiple evidence can facilitate the rapid identification of candidate drugs for COVID-19 treatment.

The emergence of SARS-CoV-2 (2019 novel coronavirus) has created the COVID-19 global pandemic. As of today (September 3, 2021), there have been over 219 million COVID-19 cases worldwide[1]. To prevent the COVID-19, several COVID-19 vaccines have received emergency approval[2], and 3 billion dosages were administered. To treat the COVID-19, many research efforts are ongoing, and the FDA approved Remdesivir[3] and Molnupiravir[4] as the COVID-19 treatments. However, none of them has proved high effectiveness for COVID-19[5,6]. Addressing the abundant needs to continue the COVID-19 drug development, many researchers have screened thousands of candidate therapeutic agents[7,8]. These agents can be divided into two broad categories: those that directly target the virus replication cycle, and those based on immunotherapy approaches either aimed to boost innate antiviral immune responses (e.g., targeting the host angiotensin-converting enzyme 2 (ACE2) that SARS-CoV-2 directly binds)[9] or to alleviate damage induced by dysregulated inflammatory responses[10]. Research on the COVID-19 therapeutic agents has created valuable knowledge and data. For example, a curated list of potential COVID-19 therapeutics is available for research, such as Comparative Toxicogenomics Database (CTDbase), which have offered valuable resources for systematic integration of accumulated COVID-19 knowledge.

Drug discovery, however, is an expensive and time-consuming process. It typically takes many years and costs billions of dollars to develop and obtain the approval of a drug. Drug repurposing is to identify existing drugs or compounds that can be efficacious to other conditions of interest. Drug repurposing via systematic integration of pharmacodynamics, in vitro drug screening, and population-scale clinical data analysis carries high potential for a novel approach by identifying highly promising drugs and their combinations to save the cost and accelerate discovery[11]. Based on accumulated genomic and pharmacological knowledge, several computational approaches have explored and identified potentially effective drug and/or vaccine candidates[12]. Examples include a network pharmacology study in protein–protein interaction (PPI) network[13], in silico protein docking[14], and sequencing analysis[15]. Another family of studies has utilized retrospective analysis of clinical data, such as electronic health records (EHRs). These studies have assessed the potential efficacy of drugs including angiotensin

[1]Center for Secure Artificial Intelligence for Healthcare, School of Biomedical Informatics, The University of Texas Health Science Center at Houston, Houston, TX, USA. [2]Research Program in Systems Oncology, Faculty of Medicine, University of Helsinki, Helsinki, Finland. [3]Center for Precision Health, School of Biomedical Informatics, The University of Texas Health Science Center at Houston, Houston, TX, USA. [4]Institute for Stroke and Cerebrovascular Disease, The University of Texas Health Science Center at Houston, Houston, TX, USA. ✉email: yejin.kim@uth.tmc.edu

nature portfolio

receptor blockers, estradiol, or antiandrogens[16]. Although network pharmacology and retrospective clinical data analysis provide complementary insight into potential drugs, few studies have integrated these complementary perspectives, particularly in COVID-19. This work attempts to identify repurposable drugs from SARS-CoV-2-drug interactions and validating the drugs from retrospective in vitro efficacy and large-scale clinical data to prioritize repurposable drugs.

In this work, we innovated the traditional network analysis by deep graph neural representation to broaden the scope from local proximity to global topology. In traditional network analysis, network proximity is defined with explicit and direct interactions[17], thus a node's local role (e.g., neighbors, edge directions) and global position (e.g., overall topology or structure) are less considered. With the recent advancement in machine learning and representation learning, the graph neural network (GNN) approach is mature for the application of its state-of-the-art technology to network pharmacology. GNN is one field of deep neural networks that derive a vectorized representation of nodes, edges, or whole graphs. The graph node embedding can preserve the node's local role and global position in the graph via iterative and nonlinear message passing and aggregation. It learns the structural properties of the neighborhood and the graph's overall topological structure[18]. Adopting GNN into the biomedical network facilitates the integration of multimodal and complex relationships. Recently GNN has shown great promise in predicting interactions (e.g., PPIs, drug-drug adverse interactions, and drug-target interactions) and discovery of new molecules[19]. GNN can also benefit drug repurposing by representing the complex interaction between drugs and diseases. A recent attempt has been made to use the GNN for drug repurposing, which builds a general biomedical knowledge graph, called Drug Repurposing Knowledge Graph (DRKG), from seven biomedical databases and utilizes the embedding to discover a therapeutic association between drugs and diseases[13]. The knowledge graph includes 15 million edges across 39 different types connecting drugs, disease, genes, and pathways from seven databases including DrugBank, Hetionet, STRING, and a text-mining-driven database. This biomedical network representation offers a general and universal understanding of the interaction between drugs, genes, and diseases.

In this study, we built the COVID-19 knowledge graph from curated COVID-19 literature, transferred the universal representation from DRKG, and then utilized deep GNN to derive repurposable drugs' representations which were rigorously validated with retrospective in vitro efficacy, reversed gene expression pattern, and large-scale EHRs (Fig. 1). Compared to the existing studies[13,17], our work's novelty can be summarized as: (i) deriving the COVID-19 knowledge representation on top of comprehensive biomedical knowledge graph, (ii) prioritizing the drug candidates based on multiple criteria including in vitro efficacy, population-based treatment effect, and reversed gene expression pattern, and iii) identifying synergistic drug combinations using complementary patterns.

## Results

### COVID-19 knowledge graph representation.

We first built a comprehensive COVID-19 knowledge graph that represents interactions between SARS-CoV-2 baits, host genes, pathways, targets, drugs (including experimental compounds), and phenotype (Fig. 1b, *Methods 1*). We then derived embedding for each drug, gene, phenotype, and SARS-CoV-2 bait using GNN. The GNN embedding method was the variational graph autoencoder with multi-relational edges (*Methods 2*)[21]. We internally validated the confidence of our knowledge graph embedding via link prediction (*Methods 3*). We compared the link prediction accuracy of our model with and without transfer learning using DRKG. Our node embedding showed high accuracy in predicting the relation in the COVID-19 knowledge graph. The initial DRKG universal embedding (without fine-tuning) achieved 0.5695 AUROC and 0.6431 AUPRC. After fine-tuning the DRKG embedding to the COVID-19 knowledge graph, we achieved AUROC 0.8121 and AUPRC 0.8524, respectively (Table S1), implying that the node embedding contains the local interaction (i.e., edges). We also visualized the node embedding using *t*-Distributed Stochastic Neighbor Embedding (*t*-SNE) (*Method 3*). We found that the node embedding of SARS-CoV-2 baits, host genes, drugs, and phenotypes were distributed separately (Fig. 2a, Fig. S2). We found that a group of antiviral and anti-inflammatory drugs (including Tenofovir, Ritonavir) was closely located to SARS-CoV-2 baits. Another group of anti-inflammatory and immunosuppressive drugs was highlighted including Cyclosporine and Dexamethasone, which were surrounded by genes related to inflammation and infection such as *CD68* and *PRDM1*. We also found a group of anticoagulants (e.g., Heparin), anti-hypertensives (e.g., Amlodipine), anti-platelet (e.g., Dipyridamole), and anti-inflammatory drugs (e.g., Indomethacin). This *t*-SNE plot showed us that our node embedding captures global topology respecting common biological knowledge.

### Initial drug ranking.

Using the rich representation of the candidate drugs, we built an initial ranking model that predicts the drug's antiviral effectiveness (*Methods 4*). The ranking model accuracy was AUROC between 0.77 and 0.90 and AUPRC between 0.17 and 0.25 (Table 1). The COVID-19 knowledge graph embedding that was boosted by general embedding from DRKG showed the highest accuracy, thanks to rich representation in DRKG. The higher AUROC/AUPRC indicated that the graph representation can encapsulate the underlying mechanism of drugs and the ranking model can pick out the drugs with potential efficacy.

### Validation with multiple sources.

From the initial drug ranking, we selected the top 300 highly-ranked drugs as potential repurposable candidates. We validated the highly-ranked drugs using a wide spectrum of validation sources such as genetic (*Methods 5*), retrospective in vivo (*Methods 6*), and epidemiological evidence (*Methods 7*), which reflects complementary aspects of drug effectiveness. Note that we did not exclude the clinical trial drugs that were used in the ranking model training.

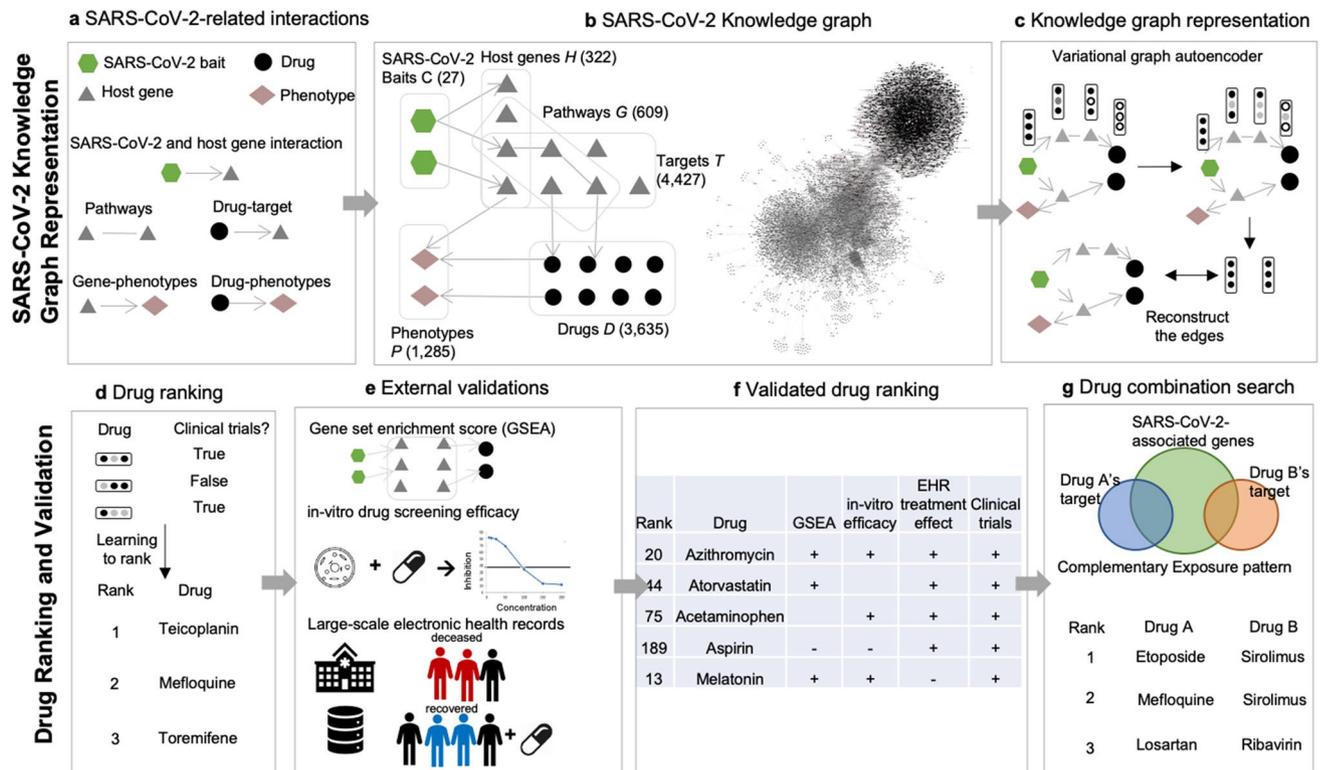

**Figure 1.** Study workflow. (**a**) We collected 27 SARS-CoV-2 baits, 322 host genes interacting with baits, 1783 host genes on 609 pathways, 3635 drugs, 4427 drugs' targets, and 1285 phenotypes, and their corresponding interactions from a curated list of the COVID-19 literature in CTDbase. (**b**) We built the COVID-19 knowledge graph with nodes (baits, host genes, drugs,targets, pathways, and phenotypes) and edges (virus–host protein–protein interaction, gene–gene in pathways, drug-target, gene-phenotype, drug-phenotype interaction). (**c**) We derived the node's embedding using the multi-relational and variational graph autoencoder[20]. We transferred extensive representation in DRKG using transfer learning. (**d**) We built a drug ranking model based on the drug's embedding as features and clinical trials as silver-standard labels. (**e**) The drug ranking was validated using drug's gene profiles, in vitro drug screening efficacy[8], and large-scale electronic health records. (**f**) We presented validated drugs with their genetic, mechanistic, and epidemiological evidence. (**g**) Using the highly ranked drug candidates, we searched for drug combinations that satisfy complementary exposure patterns[17].

## Genetic validation using gene set enrichment analysis.
For the genetic validation, we compared the gene expression signature profiles (Fig. 2b) of candidate drugs with that of SARS-CoV-2-infected host cells. We used gene set enrichment analysis (GSEA) to identify a significant association between SARS-CoV-2 and candidate drugs (*Methods 5*). As a result, we identified 183 statistically significant drugs including Gefitinib (enrichment scores or ES = − 0.70), Chlorpromazine (ES = − 0.70), Dexamethasone (ES = − 0.67), Rimexolone (ES = − 0.67), and Naltrexone (ES = − 0.64) (Fig. 2d). The lower ES scores of drugs mean the stronger signals in reversing the SARS-CoV-2 infected cell's genetic profiles. The recall and precision was 0.3, which means our prediction has moderate accuracy when compared to genetic patterns.

## Retrospective in vitro drug screening validation.
We validated the candidates by comparing them with in vitro drug screening results retrospectively. We collected four different drug screening studies that target viral entry and viral replication/infection (*Methods 6*)[7,8]. As a result, the recall was between 0.21 and 0.44 and the precision was between 0.04 and 0.18 (Table 2), implying moderate accuracy in predicting efficacy in those selected drugs. Caution is needed in interpreting the accuracy here, because the number of overlapping drugs is limited in some studies and, thus, the statistical power is limited.

## Population-based validation.
We examined drugs administered to the COVID-19 patients and estimated treatment effects of the drugs in reducing the risk of mortality among hospitalized COVID-19 patients using Optum de-identified EHR database (Table S2, Fig. S3b, *Methods 7*). The EHRs had a total of 391 drugs used for COVID-19 hospitalized patients; 138 drugs were common in the EHRs and our initial 3,635 drugs. Ten (out of 138) drugs were effective (averaged treatment effect among treated or ATT > 0 and *p*-value < 0.05) in the EHRs (Fig. 2d). Among the ten positive drugs, our method identified six positive drugs (Table 2): Acetaminophen (ATT = 0.25), Azithromycin (ATT = 0.18), Atorvastatin (ATT = 0.17), Albuterol (ATT = 0.14), Aspirin (ATT = 0.14), and Hydroxychloroquine (ATT = 0.08) (Fig. 2e) (Table S3).

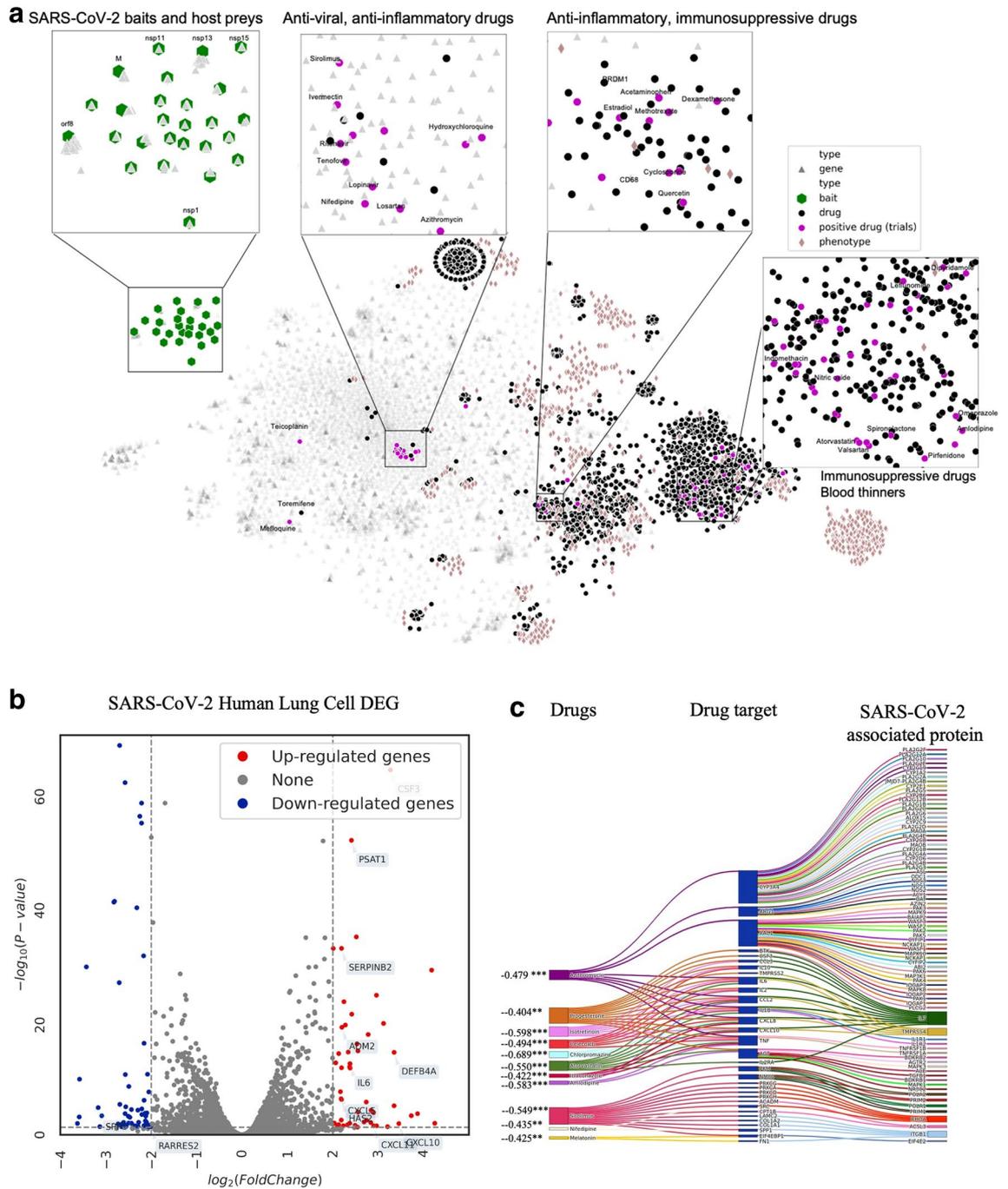

**Figure 2.** (**a**) COVID-19 knowledge graph *t*-SNE plot. Two nodes that have similar embedding are closely located in the *t*-SNE plot. We highlighted drugs undergoing clinical trials (as of July 23, 2020) to glimpse the promising repurposable drugs around the trial drugs. SARS-Cov-2 baits were the upper-left green hexagons. Genes, the gray triangles, were in the middle between baits and drugs. Drugs, the black rounds, were mixed with genes. Drugs undergoing clinical trials, the purple rounds, were closely located together. Phenotypes, the light brown diamonds, were closely located relevant genes and drugs. An interactive plot for a closer look is available in Fig. S2b–e we validated the drug ranking using four different external validation sources including. (**b**) Differentially expressed genes in SARS-CoV-2-infected human lung cells (GSE153970). (**c**) GSEA score between the infected human lung cell transcriptome and drug-induced transcriptome. (**d**) In vitro efficacy (e.g. % inhibition in viral entry and cytopathic effect assays[8]), and (**e**) treatment effects in EHRs. Figure (**c**) was created by Plotly[22] (https://plotly.com/).

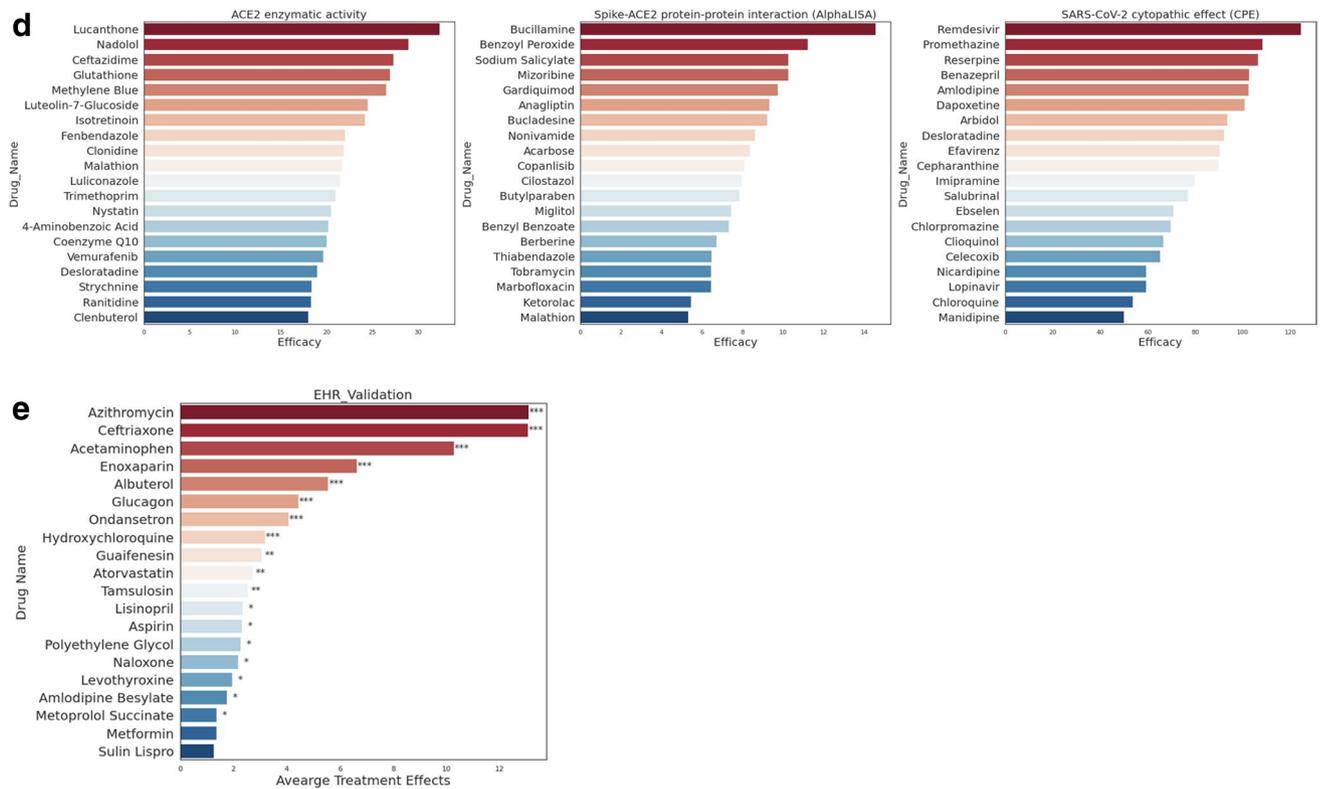

**Figure 2.** (continued)

**Validated high-ranked drugs.** Based on the extensive validation, we presented top repurposable drugs after filtering out and re-ordering the drug candidates according to the existence of validating evidence. We used a data programming technique to combine the multiple pieces of evidence (Note S3)[23]. We highlighted the most promising drugs as follows (Fig. 3a). Due to limited space, we presented the top 21 drugs in Table 3 and the remaining drugs are available in Table S4. The top 21 drugs include anti-infection, immunosuppressive or immunomodulatory, antiviral, anti-fever, antihypertensive, anti-cancer drugs, anticoagulant drugs which all have different possible functions in inhibiting SARS-CoV-2 proliferation or reducing symptoms. We highlight them in the Discussion.

**Drug combination search.** As indicated by the complexity of the COVID-19 interaction network, using single drugs to treat the viral infection might result in short term effects. To improve treatment efficacy, we further identified potential drug combinations from the top-ranking drugs with synergistic interactions without degradation in safety (*Methods 8*)[42]. We highlight the identified drug combinations as below (Table 4, Fig. 3b) and discuss potential mechanisms in discussion.

## Discussion

The objective of this study is to prioritize repurposable drugs to treat COVID-19 by a novel pipeline which harmonizes several partial pieces of evidence. In this pipeline, we applied graph neural networks to transfer a general knowledge representation from a larger knowledge network and to optimize the general knowledge representation by a human-curated COVID-19 knowledge network. The optimized COVID-19 specific knowledge representation was optimized to search and to prioritize the drugs similar to under COVID-19 trial drugs. After receiving those high-ranking drugs, we harmonize and validate those drugs' efficacy by GSEA scores, in vitro drug screening results, and population-based treatment effects. As a result, our proposed pipeline prioritized Azithromycin, Atorvastatin, Acetaminophen, and Aspirin. Also, we identified drug combinations with complementary exposure patterns: Etoposide + Sirolimus, Mefloquine + Sirolimus, Losartan + Ribavirin, and Hydroxychloroquine + Melatonin by complementary drug combination search. We highlight the identified drugs as follows:

**Antimicrobial agents.** Azithromycin and Teicoplanin can inhibit 23s ribosomes or RNA polymerase to stop the progress of infection. Some evidence supports Azithromycin regulating and/or decreasing the production of inflammatory mediators (IL-1β, IL-6, IL-8, IL-10, IL-12, and IFN-α), which might be effective to suppress viral entry[24]. Azithromycin targets ABCC1 (an inflammatory modulator) that has direct PPI with SARS-CoV-2 bait orf9c (Fig. 3a). The data imply that Azithromycin can be related to viral gene replication. In the population-based EHR validation, Azithromycin had the highest treatment effect, and it is currently under testing in a clini-cal trial(NCT04332107) to treat mild to moderate COVID-19 patients. Itraconazole can promote the production of IFN-1 that enhances viral-induced host responses[39].

**Immunosuppressive drugs.** We identified immunosuppressive drugs such as Hydroxychloroquine, Chloroquine, and Sirolimus. Hydroxychloroquine or chloroquine are anti-parasite drugs but also have effects on toll-like receptors and ACE2[31], where toll-like receptors are associated with the production of inflammatory mediators (IL-1, IL-6, TNF-α, IFN-α, and IFN-β)[45], and ACE2 is the entry receptor of SARS-CoV-2[25]. Hydroxychloroquine and chloroquine are rather controversial in terms of effectiveness[46]. Hydroxychloroquine directly targets PPT1, SIGMAR1, TRAF6, and SDC1, and it indirectly targets ECSIT and COL6A1, which had PPIs with SARS-CoV-2 baits orf8, orf9c orf10, and nsp6 (Fig. 3a). Thus, hydroxychloroquine might interfere with the SARS-CoV-2 replication. Sirolimus also works on toll-like receptors to treat COVID-19[28].

**Anti-inflammatory drugs.** Acetaminophen directly targets ACADM, CPT2, and indirectly targets ACSL3, and MARK2 which finally have PPI with SARS-CoV-2 orf9b, M, and nsp7 (Fig. 3a). This means Acetaminophen may hinder the SARS-CoV-2 assembling and replication[47]. Aspirin deactivates platelet function[48]. A recent study reports that SARS-CoV-2 may over-activate platelets and thus reduce platelet production[49]. Considering this evidence, Aspirin might be effective in COVID-19 patients by suppressing platelet function and inflammatory processes. Celecoxib is a COX2 selective inhibitor. According to a consensus docking result, Celecoxib inhibits SARS-CoV-2 main protease up to 37%[36].

**Antiviral drugs.** We identified various antiviral drugs such as Remdesivir, Lopinavir, and Tenofovir. Currently, Remdesivir has been proved to inhibit SARS-CoV-2 replication[34]. In terms of PPI between the virus bait and host prey, Lopinavir targets HMOX1, which is a host prey that binds with SARS-CoV-2 orf3a (Fig. 3a). A recent study reports that Tenofovir may prevent SARS-CoV-2 replication[41].

**Antihypertensive and lipid-lowering drugs.** We identified Atorvastatin, Amlodipine, and Nifedipine. In addition to the original function for lowering cholesterol and triglyceride levels as an HMG-CoA reductase inhibitor, Atorvastatin can treat inflammation by lowering C-reactive protein (CRP)[26]. Elevated CRP is highly associated with the aggravation of non-severe COVID-19 adult patients[50]. Also, Atorvastatin directly targets PLAT and indirectly targets HDAC2, which is a host prey of the SARS-CoV-2 nsp5. The nsp5 can assist in releasing nsp4 and nsp16, which are involved in viral replication[51]. Both Nifedipine and Amlodipine are calcium channel blockers. Nifedipine reduces the ACE2 expression[29]. In a retrospective study, Amlodipine prevents virus replication in COVID-19[52].

**Anti-cancer, antipsychotic and hormone replacement drugs.** Chlorpromazine, an antipsychotic drug shows an in vitro efficacy in inhibiting viral entry of SARS-CoV-2[38]. Progesterone decreases the severity of cytokine storms in COVID-19 patients[40]. In addition to those proposed repurposing drugs, some other highly potential drugs are also worth considering such as Bilirubin[53] and Decorin[54].

We also propose the potential drug combinations as follows and present the possible mechanism.

**Etoposide and sirolimus.** Etoposide is an anti-cancer drug that targets DNA topoisomerase 2. A recent report proposes that Etoposide can also suppress the inflammatory cytokines in COVID-19, by reducing activated cytotoxic T cells that further lead to eliminate activated macrophages[55]. There are some clinical trials to test the effectiveness of sirolimus in COVID-19 patients (NCT04341675). There is a clinical trial to test the effectiveness of combining Sirolimus, Celecoxib, and Etoposide on cancer (NCT02574728). Based on the virus bait-host prey interactome, this combination' targets interact with ten virus baits (including orf9c, orf8, orf3a, nsp1, nsp2, nsp5) without overlapping targets. We can infer this combination can be related to virus assembly in mitochondria due to an association with nsp2[51].

**Mefloquine and sirolimus.** Mefloquine not only treats malaria but also has some effects on the immune system[56]. The drug targets of Mefloquine and Sirolimus had similar baits-host prey interactome with Etoposide and Sirolimus.

**Losartan and ribavirin.** Losartan inhibits T-cell activation and also binds to ACE2[57]. Ribavirin has an ant-SARS-CoV2 function[30]. From the bait-host gene PPI, this combination's complementary drug targets had PPI with 9 virus baits including N, M, orf3a, orf8, nsp7, nsp1, nsp2, nsp13, and nsp14, which might affect the virus replication, assembling, and releasing[51].

**Hydroxychloroquine and melatonin.** Melatonin has been proposed as an adjuvant for COVID-19 treatment[58] because Melatonin can limit virus-related diseases with a high profile of safety. This might imply we can reduce the dosage of Hydroxychloroquine that decreases the risk of a long Q-T interval[31]. This speculation needs further verification.

We also observed conflicts across different validation sources. For example, Aspirin and Albuterol had positive treatment effects in EHRs validation, but there was no positive efficacy in all the four in vitro experiments. Losartan was effective in GSEA but presents negative treatment effects in EHR validation. The reason for this discrepancy might be because each validation source captures different aspects of the drug's function. The GSEA validation focused on inhibiting or activating the virus-associated host genes. The in vitro efficacy focused on viral entry, replication, or cytopathic effect. The population-based EHRs validation focused on the drugs' antiviral effect and also clinical symptom relief. For example, Acetaminophen, Azithromycin, and Albuterol are frequently given to hospitalized patients for fever, pneumonia, and shortness of breath, respectively. These drugs might not have direct effects on the virus itself. Concordance in multiple validation sources may strengthen the confidence in the drug's effectiveness. The drugs with conflicting validation results are still worth investigating.

| Embedding methods | | Ranking models | | | | |
|---|---|---|---|---|---|---|
| | Evaluation metrics | Logistic regression | Support vector machines | XGBoost | Random forest | Neural network ranking |
| COVID-19 knowledge graph embedding | AUROC | 0.6800 | 0.6915 | 0.7019 | 0.6161 | 0.7628 |
| | AUPRC | 0.0604 | 0.1149 | 0.0836 | 0.0940 | 0.1272 |
| General biomedical knowledge graph embedding from DRKG[13] | AUROC | 0.7855 | 0.8332 | 0.8500 | 0.7372 | 0.8512 |
| | AUPRC | 0.1183 | 0.1848 | 0.1439 | 0.0790 | 0.1624 |
| COVID-19 knowledge graph embedding+general embedding (proposed) | AUROC | 0.8973 | 0.7697 | 0.8934 | 0.7814 | 0.8992 |
| | AUPRC | 0.1965 | 0.1629 | 0.1701 | 0.0916 | 0.2503 |

**Table 1.** Accuracy of predicting drugs under COVID-19 clinical trials. The predictors were the drug embedding and labels that were whether a drug is under clinical trials. Logistic Regression, Support Vector Machines, XGBoost, and Random Forest were off-the-shelf models. The neural network is a customized model (*Methods*). *AUROC* area under the receiver operating curve, *AUPRC* area under the precision-recall curve.

| Validation type | Source | # overlap drugs | # true positives (TP) | # false positives (FP) | # false negatives (FN) | # true negatives (TN) | Recall TP/ (TP + FP) | Precision TP/ (TP + FN) |
|---|---|---|---|---|---|---|---|---|
| Gene profiles | GSEA scores | 580 | 55 | 128 | 128 | 269 | 0.3006 | 0.3006 |
| *In-vitro* drug screening results | ACE2 enzymatic activity[8] | 497 | 25 | 69 | 120 | 283 | 0.2660 | 0.1724 |
| | Spike-ACE2 protein–protein interaction[8] | 497 | 6 | 22 | 139 | 330 | 0.2143 | 0.0414 |
| | Cytopathic effect (NCATS)[8] | 497 | 26 | 33 | 119 | 319 | 0.4407 | 0.1793 |
| | Cytopathic effect (ReFRAME)[7] | 13 | 5 | 8 | N/A | N/A | 0.3846 | N/A |
| Population based | EHRs | 138 | 6 | 4 | 52 | 76 | 0.6 | 0.1035 |

**Table 2.** External validation of the candidate drugs using in vitro drug screening results and EHRs. *N/A* not available. False-negative or true-negative values could not be obtained because the cytopathic effect (ReFRAME) study only reports positive drugs[7]. Caution is needed in interpreting the accuracy because the number of overlapping drugs is limited in some studies and, thus, the statistical power is limited.

There are several limitations of this study. Our pipeline might have filtered out some potential drugs prematurely during the initial drug ranking step using the clinical trial drugs. The initial pool of 3635 drug candidates might miss an important set of drugs considering the fast-evolving knowledge of COVID-19 therapeutic agents.

The population-based validation was from retrospective analyses of EHRs, which are inherently incomplete and erroneous compared to randomized experimental data. Our propensity score matching and weighting approach were designed to reduce bias and confounding effects, but unmeasured or hidden confounders may exist in the EHRs data. Important laboratory values measuring the severity of COVID-19, such as White Blood Cell count, D-Dimer, and C-reactive protein, were not well documented in EHRs during the early stage of COVID-19 pandemic. The other limitation is a discrepancy between gene sets from drug-induced gene expression and SARS-CoV-2-infected cell's gene expression. cMAP provides the expression value for only 12,328 genes while the SARS-CoV-2-infected cell line (GSE153970) contains expression value for 17,899 genes. Consequently, the expression values for some genes in SARS-CoV-2 signature are missing, such as SARS-CoV-2-gp10 and SARS-CoV-2-gp01, which might cause bias. In spite of differences in cell lines as well as missing expression value of some genes, the results still have some value as a reference for further investigation.

In conclusion, this study proposes an integrative drug repurposing pipeline for the rapid identification of drugs and their combination to treat COVID-19. Our pipelines were developed from extensive SARS-CoV-2 and drug interactions, deep graph neural representation, and ranking model, and validated from genetic profiles, in vitro efficacy, and population-based treatment effects. From a translational perspective, this pipeline can provide a general network pharmacology pipeline for various diseases, which can contribute to fast drug and drug combinations repurposing.

## Materials and methods
### Building the COVID-19 knowledge graph.
To build the COVID-19 knowledge graph, we identified drug-target interactions, pathways, gene/drug-phenotype interactions from CTDbase. We collected the SARS-CoV-2 and host PPIs from a recent systematic PPI experimental study for SARS-CoV-2[51]. The graph had four

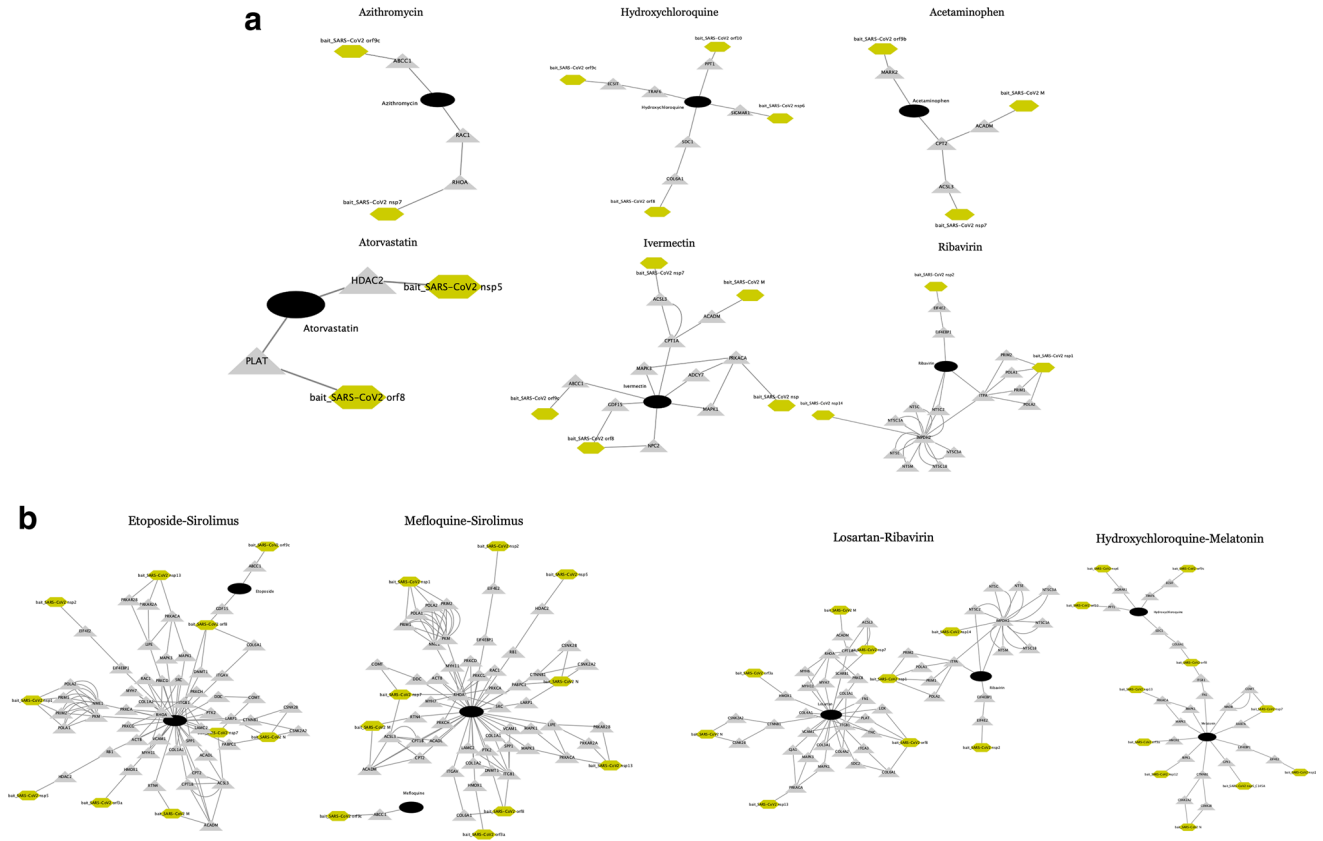

**Figure 3.** The interaction among virus baits, host preys, and drug targets. (**a**) Single drugs, (**b**) drug combinations. SARS-Cov-2 baits = green hexagons. Genes = gray triangles, Drugs = black rounds. The potentially repurposable drugs directly and indirectly target the host gene, which has PPI with the virus baits. Both figures are created by Cytoscape[94]. (https://cytoscape.org/).

types of nodes and five types of edges based on the interactions. The four types of nodes include 27 virus baits, 5677 unique host genes (from 322 host preys, 1783 genes on pathways, and 4427 drug targets, Fig. S1), 3635 drugs, and 1285 phenotypes. The five types of edges include 330 virus-host PPIs, 13,423 pairwise genes on the same pathway, 16,972 drug-target pairs, 1401 gene-phenotype pairs, and 935 drug-phenotype pairs. Details on each interaction are as follows:

*SARS-CoV-2 and human protein interactions.* We collected the SARS-CoV-2 and host interaction data from a recent work that identifies 322 high confidence PPIs between SARS-CoV-2 and the human[51]. This literature cloned 26 SARS-CoV-2 proteins in human cells and identified the human proteins that were physically associated with the SARS-CoV-2 proteins. We used the SARS-CoV-2 and human protein interaction with MiST > 0.8. In total, the virus-host interaction network consisted of 27 virus baits and 332 SARS-CoV-2-associated prey proteins.

*Drug–target interactions.* We collected drugs and targets from CTDbase's COVID-19 curated list, which contains 5065 potential targetable genes for COVID-19 with supporting biological mechanisms or therapeutic evidence. Potential compounds for SARS-CoV-2 were identified if the compounds target the SARS-CoV-2-associated genes. There were 3635 compounds that target 4427 genes. The size of the intersection between host genes interacting with baits and drug targets is 94.

*Biological pathways.* We incorporated functional pathways related to SARS-CoV-2 infection and drugs of interest. We used the Kyoto Encyclopedia of Genes and Genomes (KEGG[59]), Reactome (which were curated in CTDbase), and PharmGKB. There were 1,763 unique genes and 13,423 pairs of genes that were associated with the pathways.

*Gene/drug–phenotype interactions.* We used a curated set of phenotypes from CTDbase, which inferred the phenotypes via drug interaction and/or gene to gene ontology annotation. There were 1,285 phenotypes (i.e., biological process gene ontology) that were associated with 31 potential drugs and/or 18 SARS-CoV-2-associated genes.

| Drug name | Treated for | Targets | GSEA score | In vitro efficacy | Treatment effects in EHRs | Clinical trials | Supporting literature |
|-----------|-------------|---------|------------|-------------------|---------------------------|-----------------|-----------------------|
| Azithromycin | Anti-infection | 23 s ribosome of bacteria | + | + | + | + | 24 |
| Hydroxy-chloroquine | Immunosuppressive drug, Anti-parasite | TLR-7, TLR-9, ACE2 | NA | + | + | + | 25 |
| Atorvastatin | Lipid-lowering | HMG-CoA inhibitor | + | NA | + | + | 26 |
| Acetaminophen | Pain, fever | PGE-3, COX-1, COX-2 | NA | + | + | + | NA |
| Aspirin | Pain, fever | COX-1, COX-2 | − | − | + | + | NA |
| Albuterol | Anti-asthma | beta-2-agonist | NA | − | + | − | NA |
| Melatonin | Sleep awake cycle | Melatonin receptor | + | + | − | + | 27 |
| Sirolimus | Immunomodulatory | mTOR | + | − | NA | + | 28 |
| Nifedipine | Anti-hypertension | Calcium channel | + | + | − | + | 29 |
| Ribavirin | Anti-HCV | IMP-synthesis | NA | + | NA | + | 30 |
| Chloroquine | Immunosuppressive drug, Anti-parasite | TNF, TLR-9, ACE2 | NA | + | NA | + | 31 |
| Lopinavir | Anti-HIV | HIV-protease | NA | + | NA | + | 32 |
| Teicoplanin | Anti-infection | peptidoglycan | NA | + | NA | + | 33 |
| Remdesivir | Ebola, COVID-19 | RNA polymerase | NA | + | − | + | 34 |
| Ivermectin | Anti-parasite | Glycine receptor subu-nit alpha-3 | NA | + | NA | + | 35 |
| Amlodipine | Anti-hypertension | Calcium channel | + | + | − | + | 36 |
| Celecoxib | Anti-inflammatory | CoX2 | + | + | NA | + | 36 |
| Isotretinoin | Anti-cancer | Vitamin A derivative | + | + | NA | + | 37 |
| Chlorpromazine | Antipsychotic | D1/D2 receptor | + | + | NA | + | 38 |
| Itraconazole | Anti-fungus | Lanosterol 14-alpha demethylase | + | + | NA | + | 39 |
| Progesterone | Hormone replacement | Progesterone receptor | + | + | NA | + | 40 |
| Tenofovir | Anti-HIV | Reverse transcriptase | + | NA | NA | + | 41 |

**Table 3.** Top 22 promising drugs with supporting evidence and literature. + positive evidence, − negative evidence, *NA* not investigated. Positive in vitro efficacy if there is at least one positive efficacy in the four different in vitro experiments. Full list in Table S4.

| Drug A | Drug B | # COVID-19 genes that Drug A hits | # COVID-19 genes that Drug B hits | # COVID-19 genes that either Drug A or B hit |
|--------|--------|-----------------------------------|-----------------------------------|---------------------------------------------|
| Etoposide | Sirolimus | 2 | 22 | 24 |
| Mefloquine | Sirolimus | 1 | 22 | 23 |
| Losartan | Ribavirin | 12 | 6 | 18 |
| Hydroxychloroquine | Melatonin | 4 | 10 | 14 |
| Etoposide | Losartan | 2 | 12 | 14 |
| Acetaminophen | Chloroquine | 3 | 11 | 14 |
| Losartan | Mefloquine | 12 | 1 | 13 |
| Chloroquine | Lopinavir | 11 | 2 | 13 |
| Chloroquine | Atorvastatin | 11 | 2 | 13 |
| Acetaminophen | Melatonin | 3 | 10 | 13 |

**Table 4.** Drug combinations that satisfy the complementary exposure pattern from the top 30 drugs[43]. COVID-19 genes were defined as the host genes that have PPIs with SARS-CoV-2 baits. The full list in Table S5.

### Embedding using graph neural network.

To derive embedding from the COVID-19 knowledge graph, we utilized deep graph neural embedding with multi-relational data. We used variational graph autoencoders with GraphSAGE messages passing[18,20]. Due to uncertainty and incompleteness in our knowledge graph (i.e., COVID-19 is an emerging infectious disease and our knowledge on COVID-19 is developing), we chose to use variational autoencoders to account for the uncertainty. The graph autoencoder method is an unsupervised learning framework to encode the nodes into a latent vector (embedding) and reconstruct the given graph structure (i.e., graph adjacency matrix) with the encoded latent vector. The variational version of graph autoencoders is to learn the distribution of the graph to avoid overfitting during the reconstructing the graph adjacency matrix. In the message-passing step, each node's (entity) embedding is iteratively updated by aggregating the neighbors embedding, in which the aggregation function is a mean of the neighbor's features, concatenation with current embedding, and a single layer of a neural network on the concatenated one. We set different weight matrices for each of the five types of edges. Since our objective is to use the drug embedding to discover drugs that can functionally target SARS-CoV-2-associated host genes, the model was trained to reconstruct the missing interaction using the node embeddings as an unsupervised manner. We set the embedding size as 128 after several trials. We used PyTorch Geometric for implementation. The model structure was $(1 \times 400) \rightarrow$ Graph convolution to $(1 \times 256) \rightarrow$ RELU $\rightarrow$ Dropout $\rightarrow$ Concatenation of multiple edge types $\rightarrow$ Batch norm $\rightarrow$ Graph convolution to $1 \times 128$ (mean) and $1 \times 128$ (variance).

We further boosted the representativeness of the embedding by transferring DRKG universal embedding to our embedding. The DRKG embedding contains general biological knowledge (e.g., drug embedding was derived from molecular structures, targets, anatomical therapeutic chemical classifications, side effects, pharmacologic classes, and treating diseases)[13]. By transferring the rich representation of DRKG to the COVID-19 knowledge graph, we can derive embeddings that are more faithful to underlying pharmacokinetics and pharmacodynamics. To this end, we initialized the COVID-19 knowledge graph node embedding with DRKG embedding and fine-tuned the node embedding by updating them via GNN's message passing and aggregation. (Note S1).

**Evaluating the knowledge graph embedding.** We internally validated the confidence of our knowledge graph embedding via link prediction to confirm if the node embedding can capture the network topology centered by SARS-CoV-2. We measured an accuracy to predict interactions between the nodes (SARS-CoV-2 baits, genes, drugs, and phenotypes). We randomly selected 10% of the edges for validation.

We also visualize the node embedding using lower-dimensional projection to observe the distribution of high-dimensional node embedding. The $t$-SNE plot projects a high-dimensional vector into a low-dimension vector while preserving the pairwise similarity between nodes, thus allowing us to examine the high-dimensional node embedding with low-dimension (e.g., 2-dimensions) visualization.

**Initial drug ranking.** After we derived the drug embedding, we built a ranking model to select the most potent drugs. We hypothesized that, because drugs testing in clinical trials are potentially efficacious in treating COVID-19, a drug that is similar to these trial drugs can have potential efficacy too. This drug ranking was an initial filtering step to select possibly potent drugs out of 3,635 candidates. The drugs under clinical trials were extracted from NIH ClinicalTrials.gov's interventional trials. 99 trial drugs were matched to the CTDbase's 3635 drugs. The remaining drugs without matched clinical trials were regarded as having negative efficacy. We designed a customized neural network ranking model based on Bayesian pairwise ranking loss[60]. The architecture was two fully connected layers (with the size of $128 \rightarrow 128 \rightarrow 1$) with residual connection, nonlinear activation (ReLU), dropout, batch norm in the middle, and the optimization loss (Bayesian pairwise ranking loss). Baseline ranking models to compare were logistic regression, support vector machine, XGBoost, and Random forest.

We measured the accuracy of the drug ranking model using the area under the receiver operating curve (AUROC) and area under the precision-recall curve (AUPRC) with 50% training and 50% test cross-validation. We purposely set the portion of the training set lower because the clinical trials are not our sole "gold standard" to prioritize drugs. Note that the unsupervised knowledge graph embedding and the supervised drug ranking were independent. We tried to avoid using the supervised label (clinical trials drugs) in the knowledge graph embedding because the drugs being considered in clinical trials do not guarantee the efficacy of the drugs.

**Genetic validation.** We obtained the gene expression signature of SARS-CoV-2 from SARS-CoV-2 infected human lung cells[61], and obtained the drug's gene expression signature profile from the Connectivity Map (cMAP) database (GSE92742 and GSE70138)[62]. We determined whether the drug's gene expression signature is negatively correlated with that of SARS-CoV-2 based on the enrichment score (ES)[63]. The combining ES < 0 and $p$-value < 0.05 was considered as the threshold to determine that a drug has a complementary expression pattern with COVID-19 infections (Note S3).

**Retrospective in vitro drug screening validation.** We validated the highly ranked candidate drugs by retrospectively comparing them with efficacious drugs in multiple in vitro drug screening studies. We utilized four drug screening studies targeting viral entry (ACE2 enzymatic activity, Spike-ACE2 protein–protein interaction) and viral replication/infection (cytopathic effect), which are obtained from NCATS OpenData COVID-19 Portal and Riva et al. study[7,8]. The two viral entry assay studies screened 2,678 compounds in the NCATS Pharmaceutical Collection and 739 compounds in the NCATS Anti-infectives Collection[64]. In the viral entry assay, a drug was regarded as efficacious if efficacy value was larger than 10 and 0 for ACE2 enzymatic activity and Spike-ACE2 interaction, respectively (the efficacy value was defined as an % inhibition at infinite concentration subtracted by % inhibition at zero concentration by curve fitting). The two cytopathic effects studies use either the NCATS collections or the ReFRAME drug library on the same Vero E6 cell[65]. In the NCATS cytopathic effect study, a drug was regarded as efficacious if the efficacy value was larger than 10. In the ReFRAME study, a drug was regarded as efficacious if the drug inhibited infection by 40% or more[7]. We calculated precision and recall between the predicted (top 300 highly-ranked) drugs and the efficacious drugs in each screening result (Fig. 2c). We focused on only those drug candidates that are included in the compound library in the screening study.

**Population-based validation.** We investigated drugs administered to the COVID-19 patients and estimated treatment effect using counterfactual analysis. We used Optum® de-identified EHR database (2007–2020), which is (non-experimental data, as opposed to randomized clinical trials).

In 140,016 positive COVID-19 patients, there were a total of 34,043 hospitalized COVID-19 patients; we selected 3200 deceased patients during the hospitalization and 15,078 recovered patients with medication history and length of stay > 2 days.

The key to estimate treatment effect is to reduce bias or confounders in EHRs to control the difference of confounding variables between those who received and did not receive treatment. We calculated the average treatment effect on the treated (ATT) by using propensity score matching (PSM) and weighting to build the cohort (Note S2). From the selected hospitalized patients, we built a cohort with 2827 cases (deceased) and 2774 controls (recovered) that follow similar distributions in terms of demographics (race, ethnicity, sex, age) and admission severity (body temperature and $SPO_2$) using PSM. The time period of severity risk factors was from before 2 h of admission and to after 6 h of admission. After we derived the matched cohort, there were a total of 391 medications that were administered in at least 35 patients. We calculated the treatment effect of the 391 medications using the average treatment effect among treated or ATT. For the inverse propensity score weighting, we considered demographics (age, gender, race), admission conditions (body temperature, $SPO_2$), comorbidities (cancer, chronic kidney disease, obesity, a serious heart condition, solid organ transplant, COPD, type II diabetes, and sickle cell disease), and drug history before the treatment of interest. We assumed a drug is effective if ATT > 0 and the $p$-value is < 0.05. A full list of the drug's ATT coefficient is in Table S3. The Optum de-identified EHR database within this study has been approved by the Committee for the Protection of Human Subjects (UTHSC-H IRB) under protocol HSC-SBMI-13-0549.

**Drug combination search.** We identified efficacious drug combinations from top-ranked drugs. Our approach is to leverage drug targets and COVID-19 associated host genes. Our working hypothesis was based on the Complementary Exposure pattern that "a drug combination is therapeutically synergistic if the targets of the individual drugs hit the disease module, but target a separate neighborhood"[43]. We searched the drug combinations within the top 30 drugs. We identified the COVID-19 modules from human protein interactomes that are physically associated with SARS-CoV-2 baits[51]. The drug's targets were identified from CTDbase's COVID-19 curated list. We counted the number of genes in the COVID-19 module that a drug combination hits, where the drug combination's targets are disjoint.

## Data availability
Code is available at https://github.com/yejinjkim/drug-repurposing-graph. Data is available at Supplementary tables. The raw COVID-19 knowledge graph data derived from CTDbase (http://ctdbase.org/).

## Acknowledgements

We thank Lu Chen and Hall Matthew from NCATS for in vitro efficacy experiments.

## Author contributions

X.J., J.T. and Z.Z. provided motivation for this study; X.J., K.H., and Y.W. collected necessary data; Y.K. built graph representation and ranking models; X.J. and L.C. performed population-based validation; Y.W. and K.H. performed genetic validation; K.H. and J.T. derived drug combinations; K.H. prepared plots; Y.K., K.H., Y.W., J.T., Z.Z., S.S., and X.J. wrote manuscript; and K.H. and S.S. provided clinical interpretation.

## Funding

Y. K. was supported in part by CPRIT RR180012, R01AG066749, and R01AG066749-01S1. K. H. was supported by CPRIT RR180012. Y. W. and J. T. were supported by ERC starting Grant No. 716063 and Academy of Finland funding No. 3176880. Z. Z. is partially supported by NIH grant R01LM012806 and CPRIT grant CPRIT RP180734. X. J. is CPRIT Scholar in Cancer Research (RR180012), and he was supported in part by Christopher Sarofim Family Professorship, UT Stars award, UTHealth startup, the National Institute of Health (NIH) under Award Number R01AG066749 and R01AG066749-01S1.

## Competing interests

The authors declare no competing interests.

## Additional information

**Supplementary Information** The online version contains supplementary material available at https://doi.org/10.1038/s41598-021-02353-5.

**Correspondence** and requests for materials should be addressed to Y.K.

**Reprints and permissions information** is available at www.nature.com/reprints.

**Publisher's note** Springer Nature remains neutral with regard to jurisdictional claims in published maps and institutional affiliations.